\documentclass[11pt]{article}
\usepackage{float}
\usepackage[inline,shortlabels]{enumitem}
\usepackage{amssymb, amsmath, amsthm}
\usepackage{mathrsfs}
\usepackage{tabulary}
\usepackage{rotating}
\usepackage{enumerate}
\usepackage{setspace}
\usepackage{multirow}
\usepackage{paralist}
\usepackage{listings}
\usepackage[T1]{fontenc}
\usepackage[utf8]{inputenc}
\usepackage[english]{babel}
\usepackage{dsfont}
\usepackage{hyperref}
\usepackage{subcaption}
\usepackage{amsmath}
\usepackage{graphicx}
\usepackage{url}
\usepackage{booktabs}

\usepackage{authblk}
\usepackage{etoolbox}
\usepackage{longtable}
\usepackage[super,sort&compress]{natbib}
\setcitestyle{comma,numbers,super,open={},close={}} 

\usepackage{color}
\usepackage{bibentry}

\newcommand\BibTeX{{\rmfamily B\kern-.05em \textsc{i\kern-.025em b}\kern-.08em
T\kern-.1667em\lower.7ex\hbox{E}\kern-.125emX}}

\usepackage{caption}
\usepackage{hyperref}
\usepackage{subcaption}

\usepackage{censor}

\usepackage[hmargin=2.54cm,vmargin=2.54cm]{geometry}

\newcommand{\E}{\mathsf{E}}

\usepackage{pdflscape}
\usepackage{amsmath, booktabs, caption, longtable}
\usepackage{tabularx} 
\usepackage{threeparttable}
\usepackage{tablefootnote}

\usepackage[hmargin=2.54cm,vmargin=2.54cm]{geometry}

\title{Associations between pain-management treatments and opioid use disorder risk among Medicaid patients}
\vspace{3cm}

\author[1]{Kara E. Rudolph, PhD} 
\author[1]{Nicholas T. Williams, MPH}
\author[2]{Ivan Diaz, PhD}
\author[1]{Sarah Forrest, MPH}
\author[1]{Katherine L. Hoffman, MPH}
\author[3]{Hillary Samples, PhD}
\author[1]{Mark Olfson, MD}
\author[2]{Lisa Doan, MD}
\author[2]{Magdalena Cerda, DrPH}
\author[1]{Rachael Ross, PhD}

\affil[1]{\footnotesize Department of Epidemiology, Mailman School of Public Health, Columbia University}
\affil[2]{\footnotesize New York University Grossman School of Medicine.}
\affil[3]{\footnotesize Rutgers Institute for Health, Rutgers University}

\date{}
\begin{document}

\maketitle









\begin{abstract}
\noindent \textbf{Introduction:} Chronic pain patients are at increased risk of opioid-misuse. Less is known about the unique risk conferred by each pain-management treatment, as treatments are typically implemented together, confounding their independent effects. 
We estimated the extent to which pain-management strategies were associated with risk of incident opioid use disorder (OUD) for those with chronic pain, controlling for baseline demographic and clinical confounding variables and holding other pain-management treatments at their observed levels. \\

\noindent \textbf{Methods:} We used data from two chronic pain subgroups within a cohort of non-pregnant Medicaid patients aged 35-64 years, 2016-2019, from 25 states: 
1) those with a chronic pain condition co-morbid with physical disability (N=6,133) or 2) those with chronic pain without disability (N=67,438). We considered
9 pain-management treatments: prescription opioid i) dose and ii) duration; iii) number of opioid prescribers; opioid co-prescription with iv) benzodiazepines, v) muscle relaxants, and vi) gabapentinoids; vii) non-opioid pain prescription, viii) physical therapy, and ix) other pain treatment modality. 
Our outcome was incident OUD.\\

\noindent \textbf{Results:} 
Having an opioid and gabapentin co-prescription or an opioid and benzodiazepine co-prescription was statistically significantly associated with a 16-46\% increased risk of OUD. Opioid dose and duration also were significantly associated with increased risk of OUD. Physical therapy was significantly associated with an 11\% \textit{decreased} risk of OUD in the subgroup with chronic pain but no disability.\\

\noindent \textbf{Conclusions:} Co-prescription of opioids with either gabapentin or benzodiazepines may substantially increase risk of OUD. 
More positively, physical therapy may be a relatively accessible and safe
pain-management strategy.

 \end{abstract}

\newpage

\doublespacing

\section*{Introduction}
Opioid overdoses continue to rise in the United States (US) to levels termed ``staggering'' \cite{nyt}. 
Mortality rates involving opioids have nearly quadrupled over the past 10 years \cite{nida2023}, underscoring the urgency of accelerating opioid misuse prevention and treatment efforts \cite{cms}. A large body of research indicates that individuals with non-cancer chronic pain are at particular risk of opioid misuse \cite{mikosz2020indication,marshall2019considerations,dunn2010opioid,orhurhu2019trends,volkow2016opioid,cerda2021critical}, and the subset of chronic pain patients that also have a physical disability (termed ``high-impact chronic pain" \cite{pitcher2019prevalence,interagency_pain_research_coordinating_committee_national_2016}) may be at even greater risk 
 \cite{hoffman2023independent}. 

However, identifying subgroups at increased risk of adverse opioid-related outcomes is of limited use in the absence of evidence-based strategies that could reduce their risk. 
Given the centrality of pain management 
for those with a chronic pain condition, identifying treatments that are particularly risky or protective and amenable to intervention would provide evidence-based targets for prevention efforts. 
Opioid prescribing continues to be one of the most common pain-management treatments \cite{mikosz2020indication}, but 
 opioid prescriptions at higher doses and longer durations, use of multiple opioid prescribers, and co-prescriptions of opioids with medications that have respiratory-depressive effects, like benzodiazepines, muscle relaxants, and gabapentin/pregabalin, have been implicated in increasing risk of opioid misuse \cite{savych2019opioids, edlund_role_2013, glanz2019association,peirce_doctor_2012,cho2020risk,volkow2016opioid,rose2018potentially, gressler2018relationship,ford2018disability,sun2017association,li2020risk,khan2022comparative,gomes2017gabapentin,kuehn2022growing,corriere2023concurrent,peckham2018all,olopoenia2022adverse,bykov2020association}.
Currently, a multi-modal approach to pain treatment is recommended 
that consists of $\ge 2$ of the following: i) medications (opioid and non-opioid), ii) restorative therapies (e.g., physical therapy, massage therapy), 
iii) interventional procedures (e.g., epidural steroid injections, nerve blocks), 
iv) behavioral health (e.g., counseling), and v) complementary/integrative health (e.g., acupuncture, yoga) \cite{interagency_pain_research_coordinating_committee_national_2016}. By including non-opioid treatments in pain management, reliance on opioid-based treatments may be reduced.  Although 
these 
recommended non-opioid modalities may effectively treat pain \cite{interagency_pain_research_coordinating_committee_national_2016}, with the exception of counseling, whether or not they reduce opioid misuse remains an open question 
\cite{
vowles2014acceptance,aytur2021neural,roberts2022mindfulness,thomas2017mindfulness,priddy2018dispositional,xie2021web,craft2020episodic}. 

In studying the relationship between pain-management treatments and opioid misuse outcomes, previous research considered one or a select few pain-management treatments at a time, leaving others unmeasured and unaccounted for. However, chronic pain patients may experience multiple treatments concurrently or close together in time. Consequently, this previous research was likely unable to isolate 
the unique contribution of each treatment, 
as the treatment being considered was conflated with other, unmeasured treatments. We sought to address this gap by considering a larger number of pain-management treatments across the multiple modalities specified above, using a novel approach to isolate the extent to which each was associated with an 
increased or decreased risk of incident opioid use disorder (OUD) for those with i) a chronic pain condition co-morbid with physical disability or ii) a chronic pain condition without physical disability. We hypothesized that opioid-involved pain-management strategies would be associated with 
more risk and that non-opioid strategies would be associated with reduced risk. We estimated the risk associated with each treatment 
among adult, non-dual-eligible Medicaid patients without baseline OUD, 2016-2019. 

\section*{Methods}
The study was approved by the Columbia University Institutional Review Board. All code to replicate cohort and variable creation and statistical analyses is given in the GitHub repository: \url{blinded for review}. Additional methodological details are in Section S1 of the appendix.

\subsection*{Data and Cohort}
\begin{figure}[H] 
\captionsetup{justification=raggedright,singlelinecheck=false}
  \caption{Study measure characterization timeline}
\centering
\includegraphics[width=0.99\textwidth]{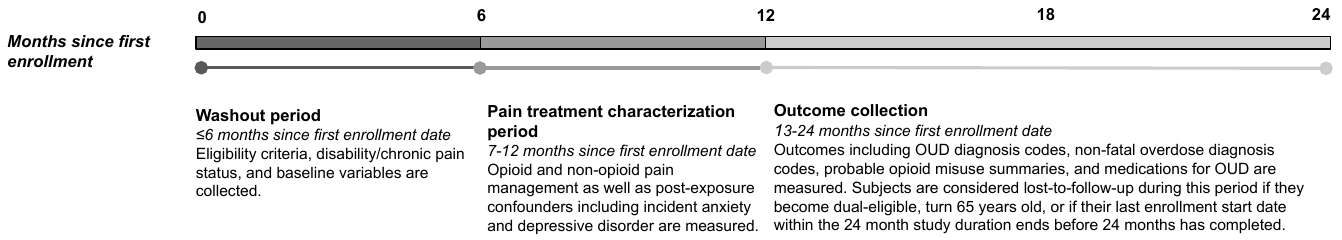}
\label{fig:timeline}
\end{figure}

We utilized data from Medicaid T-MSIS Analytic Files (TAF): Demographics, Other Services, Inpatient, and Pharmacy claims, 2016-2019, and included non-pregnant Medicaid patients aged 35-64 years in 25 states that had implemented Medicaid expansion under the Affordable Care Act by 2014. 
Because individuals receiving Social Security Disability Insurance  transition to Medicare after 24 months \cite{rupp2012longitudinal}, we included claims for a maximum of 24 months post-enrollment. We permitted enrollment on or after January 2, 2016 and 
required 12 months of continuous enrollment. 
During the first 6 months of enrollment (months 1-6), we evaluated eligibility criteria, characterized baseline covariates, and determined membership in one of the two chronic condition subgroups described below (if any). We measured the set of pain-management treatment variables in months 7-12. We measured outcomes during months 13-24 in the primary analysis and months 13-18 in a secondary analysis. Figure \ref{fig:timeline} depicts this timeline. 

Figure S1 illustrates the exclusion/inclusion criteria. Using the initial 6-month period, we excluded those dual-eligible for Medicare, lacking a Medicaid eligibility code or with undetermined disability status, or who had a claim indicating OUD during this period. 

Patients were censored at the point they became Medicare-eligible, 24 months-post enrollment, at the end of 2019, or if they had disenrolled by the end of the study period.  

\subsection*{Measures}

\subsubsection*{Covariates}

Baseline covariates included: age in years, sex, race/ethnicity, primary language (English), marriage/partnership status, household size, veteran status, and 
whether the beneficiary's income was likely $>$138\% of the Federal Poverty Level; and clinical measures, including any inpatient or outpatient diagnosis of bipolar disorder, anxiety disorder, attention deficit hyperactivity disorder (ADHD), depressive disorder, or other mental disorder  \cite{samples2018risk}, and whether or not the beneficiary had a claim for mental health counseling. In addition to baseline covariates, we controlled for anxiety and depressive disorder diagnoses occurring in months 7-12, as these conditions have been shown to be affected by pain and also increase risk of OUD \cite{cohen2021chronic, fox2014disability, whitney2018cardiometabolic,mills2019chronic, marshall2019considerations,sullivan2010risks, segrin2018indirect,cance2021examining,zoorob2017bowling, mclean2016there, monnat2018factors,krueger2017have,dasgupta2018opioid,ledingham2022perspectives}. We also controlled for presence of a claim for mental health counseling during months 7-12, as this could be a marker for a mental health condition that was otherwise not observed in the claims diagnostic codes. 

\subsubsection*{Chronic condition subgroups}

Using claims from months 1-6, we determined whether each patient met criteria for either: (1) chronic pain and co-occurring physical disability (CPPD), or (2) chronic pain only (CP). We identified chronic pain status by non-cancer diagnoses (ICD-10 codes) typically associated with chronic pain \cite{mannes2023increasing}, occurring at least two times 90 days apart for the same region of the body \cite{miller2019prevalence}. We identified a physical disability by including eligibility codes indicating eligibility based on disability, while excluding codes indicating common disabilities which were unlikely to result in periods of acute or chronic pain, e.g. developmental disabilities, hearing or sight impairments) \cite{hoffman2023independent}. 

\subsubsection*{Exposures}

We considered the following variables related to the prescribing of opioids for pain (excluding opioids used for OUD treatment): 1) maximum daily opioid dose in morphine milligram equivalents (MME) considering doses of all pain-related fills, 
2) proportion of days covered (out of the 6 month period) for these opioid prescriptions, 3) number of unique opioid prescribers, and co-prescription ($>$25\% overlap, first prescription $\ge5$ days supply) 
of opioids for pain with 4) benzodiazepines, 5) muscle relaxants, and 6) gabapentinoids. We also considered the following non-opioid pain management variables: 7) presence of a non-opioid pain prescription (noting that claims will poorly capture over-the-counter medications), 8) physical therapy, and 9) other pain treatment modalities comprising at least one of the following: a) ablative techniques, b) acupuncture, c) joint and nerve blocks, d) botulinum toxin injections, e) chiropractic work, f) transcutaneous electrical nerve stimulation, g) epidural steroids, h) intrathecal drug therapies, i) massage therapy, and j) trigger point injection.  

\subsubsection*{Outcomes}
The primary outcome was new OUD diagnosis (which we refer to as "incident OUD", because those with OUD during the washout period were excluded), defined as the presence of any of the following: ICD-10 diagnosis codes indicating opioid abuse or dependence; nonfatal, unintentional opioid overdose ICD-10 diagnosis codes; and 
medication for OUD treatment (MOUD; methadone, buprenorphine, or injection naltrexone) \cite{cochran2017examination}. 
As a secondary outcome, we defined incident OUD diagnosis using only ICD-10 diagnosis codes indicating 
opioid abuse or dependence 
\cite{samples2018risk, samples2022psychosocial}. 
 %
 For the primary analysis, outcomes were assessed months 13-24 post-enrollment, and a censoring indicator distinguished those who remained enrolled versus did not. However, it is plausible that our censoring model may be poorly estimated, so to reduce reliance on it, we examined only a 18-month study period in the secondary analysis. 

\subsection*{Statistical Analysis}
We estimated the association between each pain-management treatment of interest and incident OUD risk, adjusting for covariates, holding other pain-management treatments at their observed levels, 
and stratifying by chronic condition subgroup. We formally define these quantities and the assumptions under which they can be interpreted as effects in Section S1.3 of the appendix. 
For binary pain-management treatments (e.g., presence of opioid co-prescription with gabapentin), 
we estimated the expected difference in OUD risk if everyone in the subgroup received the treatment versus no intervention on treatment. 
For non-binary pain-management treatments (e.g., maximum daily opioid dose), 
we estimated the expected difference in OUD risk if everyone in the subgroup had their value increase by 20\% (i.e., multiplied each person's value by 1.2) versus no intervention on treatment. 

The above statistical estimands are considered modified treatment policies and are especially well-suited for questions that consider a set of interrelated exposure or treatment variables where interventions can be considered across the set of variables \cite{haneuse2013estimation,munoz2012population,young2014identification,diaz2023nonparametric}. 
We estimated these associations using a doubly robust, nonparametric targeted minimum loss-based estimator \cite{diaz2023nonparametric}. We used a cross-fitted version of this estimator with 2-folds for the CP subgroup and 5-folds for the CPPD subgroup. This estimator fits regressions for the outcome mechanism, treatment mechanism, and censoring mechanism. These regressions were fit using an ensemble of machine learning algorithms\cite{vanderLaanPolleyHubbard07} consisting of intercept-only models, main-terms generalized linear models, gradient boosted machines\cite{xgboost}, multivariate adaptive regression splines\cite{earth}, neural networks\cite{nnet}, and random forests\cite{ranger}.  

Code for all data cleaning and statistical analyses is available at \url{blinded for review}. 

\section*{Results}
There were $N=6,133$ Medicaid patients in the CPPD subgroup and  $N=67,438$ in the CP subgroup. Table \ref{tab:desc} gives the descriptive statistics of these subgroups, which 
were similar demographically but 
with generally higher prevalence of mood disorders in the CPPD subgroup. 
Over half of patients (61\%) in the CPPD group received an opioid pain prescription in the 7-12 month period; 
opioid prescription rates were lower but still common in the CP subgroup at 40\%. Co-prescriptions of opioids with benzodiazepines, gabapentinoids, and muscle relaxants were common in both subgroups. Of the non-opioid pain-management treatments, 
over one-quarter of those in either group 
had a claim for physical therapy in the 7-12 month period. 

[Table \ref{tab:desc} here]

\begin{figure}[ht]
\caption{Estimated effects of each pain-management treatment on incident OUD and 95\% CIs for the cohort with (A) co-occurring chronic pain and disability, and (B) chronic pain alone.}
\centering
 \vspace{-.3cm}
 \subfloat[Co-occurring chronic pain and disability \label{fig:CPPD}]{
\includegraphics[width=\textwidth]{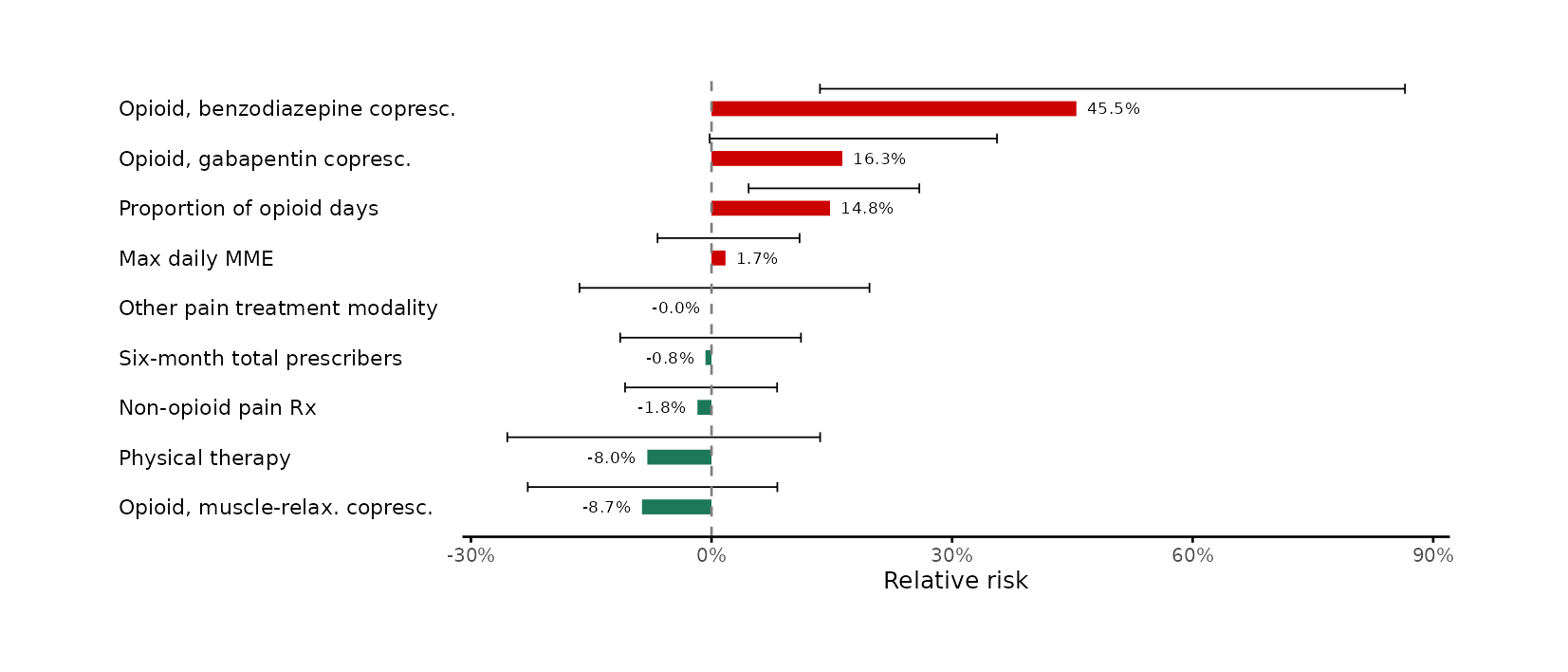}
 }
 \\
 \vspace{-1cm}
  \subfloat[Chronic pain alone \label{fig:cp}]{
\includegraphics[width=\textwidth]{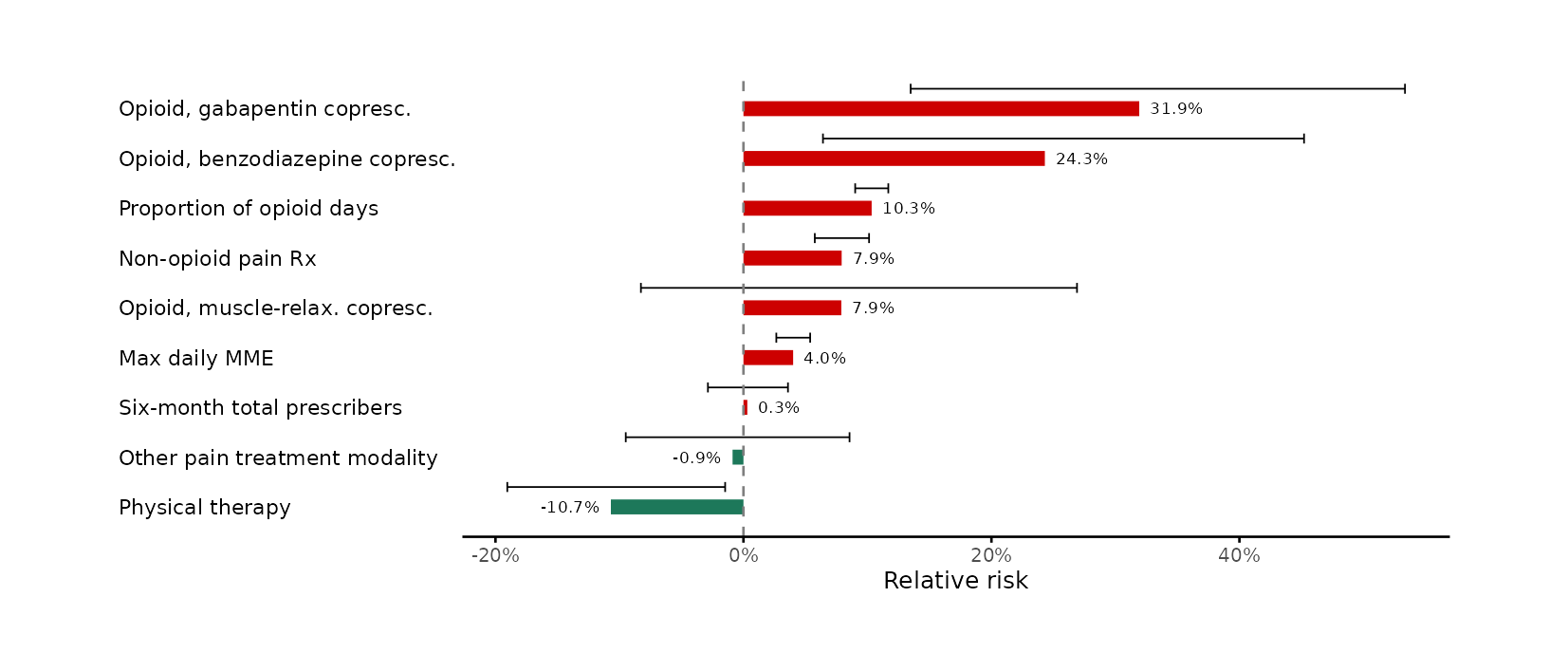}
 }
   \label{fig:res}
   
\end{figure}

Figure \ref{fig:res} shows the estimated relative risks (as a percent, e.g., 1.2 is 20\% and 0.8 is -20\%) of each pain-management treatment on incident OUD by month 18 of follow up (month 24 of cohort enrollment) and their 95\% confidence intervals, controlling for covariates and holding the other pain-management treatments at their observed values. 

Several pain treatments were associated with increased OUD risk across both subgroups. Having an opioid and gabapentin co-prescription was statistically significantly associated with a 32\% (95\% CI: 13-53\%) and 16\% (95\% CI: 0-36\%) increased risk of OUD for CP and CPPD subgroups, respectively. 
Given that average 12-month risk of OUD is relatively high in these groups---developing in 10\% of CPPD patients and 4\% of CP patients (Table 1)---these relative risk estimates correspond to appreciable absolute risk. 
Opioid and benzodiazepine co-prescription was also associated with substantial increased OUD risk (24\%, 95\% CI: 6-45\% in the CP group; 45\%, 95\% CI: 14-86\% in the CPPD group). In addition, increases in the proportion of days covered by an opioid was also statistically significantly associated with increased risk of OUD in both groups and increases in opioid dose was associated with increased risk in the CP group. For example, for CP, increasing the proportion of days covered by 20\% was associated with a 10\% (95\% CI: 9-12\%) increased risk of OUD, and increasing the maximum daily dose by 20\% was associated with a 4\% (95\% CI: 3-5\%) increased OUD risk. 

On the other hand, physical therapy was associated with a \textit{decreased} risk of developing OUD; in the CP subgroup, this association was a 
statistically significant 11\% decreased risk (95\% CI: 1-19\%). 

[Figure \ref{fig:res} here] 

We repeated the above analyses defining OUD based only on ICD-10 codes (Figure S2); results were largely unchanged. We also repeated the above analyses measuring incident OUD after only 6 months of follow-up (Figure S3); results were largely unchanged.

\section*{Discussion}
In a large, carefully constructed cohort of over 70,000 adult, non-elderly Medicaid patients with chronic pain, we  
found 
robust evidence of which treatments were associated with 
the most risk of developing OUD and which were associated with reduced risk. 
In particular, our findings caution against the co-prescription of opioids and gabapentin, which has become a mainstay of pain treatment, 
prescribed for an estimated 20\% of pain patients, but which we found was associated with increased risk of incident OUD by up to 32\%. 
On a positive note, our findings highlight the potential benefits of physical therapy for those with chronic pain, as it was associated with an estimated 11\% 
 statistically significant decreased risk of OUD. While previous literature found evidence that physical therapy reduced risk of being prescribed an opioid \cite{george2020physical}, ours is the first to show that physical therapy---even when used in conjunction with prescribed opioids---may reduce risk of an opioid misuse outcome.

In terms of identifying risky pain treatment strategies, our findings largely corroborated previous research; our unique contributions were in terms of our careful quantification of such associated risk, holding all other treatments constant at their observed levels. A large body of previous research established that higher doses of opioids and opioids taken for longer duration each conferred increased risk of opioid misuse, including the development of OUD \cite{savych2019opioids, edlund_role_2013, glanz2019association,rose2018potentially,cho2020risk,volkow2016opioid}. We too found that these aspects of opioid prescribing were risky---a 20\% increase in maximum daily opioid dose was associated with a 2-4\% increase in OUD risk (which was statistically significant for the larger CP subgroup), and a 20\% increase in proportion of days covered by an opioid prescription was associated with a 10-15\% increase in OUD risk, which was statistically significant in both subgroups. Also aligned with previous research \cite{rose2018potentially, gressler2018relationship,cho2020risk,ford2018disability,sun2017association,li2020risk,khan2022comparative,gomes2017gabapentin,kuehn2022growing,corriere2023concurrent,peckham2018all,olopoenia2022adverse,bykov2020association}, we found co-prescribing of opioids with the respiratory depressive medications gabapentin and benzodiazepines 
to be risky, associated with a statistically significant 
16\%-46\% increased OUD risk, depending on medication and subgroup. 

However, we did not find significant associations between OUD and other aspects of treatment that previous research found to be risky. For example, in contrast with previous research \cite{peirce_doctor_2012,rose2018potentially}, we did not estimate a statistically significant association between the total number of opioid prescribers and OUD. This discrepancy could be because after holding opioid dose, duration, and co-prescriptions constant (which previous research did not do), there is little additional risk conferred by the total number of prescribers. Similarly, we did not estimate a statistically significant association between the co-prescription of opioid and muscle relaxants and OUD; previous research found evidence that their co-prescription increased risk of opioid overdose \cite{li2020risk,khan2022comparative}, but was again limited by not conditioning on other common aspects of pain treatment (thus, their associations could have been confounded). 
Finally, %
although we hypothesized that other pain treatment modality would be associated with decreased risk of OUD, we instead found no significant association, 
perhaps because our definition comprised a set of treatments that were 
disparate. 

This study improved upon previous research in several ways. First, we measured a larger number of pain-management treatments that may occur concurrently, at least in part, and 
used a novel statistical approach to 
better isolate the unique contribution of each treatment to OUD risk. 
Second, we used a large, policy-relevant cohort of Medicaid patients. An estimated 40\% of people with OUD are covered by Medicaid \cite{KaiserOpioid}; thus, 
this population has the potential to be particularly impacted by policy change. Third, we used a flexible, robust nonparametric estimator to make our analysis as assumption-lean as possible. For example, unlike previous analyses attempting to estimate risk of pain treatments on OUD and other opioid misuse outcomes, 
we do not assume that we can correctly specify a parametric model, which would be 
a tenuous assumption given the complex web of time-varying demographic, clinical, provider, and contextual factors 
that contribute to 
risk of such outcomes. 
Our estimator 
is also doubly robust, meaning that as long as we can estimate either the outcome model consistently or the treatment and censoring models consistently, then our estimate is expected to be unbiased. 

This study is nonetheless limited in several respects. 
In claims data, 
 information (e.g., diagnoses) may not be recorded at all, may be recorded inaccurately, or may be recorded at an inaccurate time. One relevant example is the challenge of measuring  OUD in claims; previous research found that using ICD-10 diagnosis codes in Medicaid claims data captured 84\% of OUD cases identified by record reviews by addiction medicine clinicians 
 \cite{mcneely2020sensitivity}. Consequently, for our primary analysis, we used a 
 more expansive definition of OUD (including opioid poisonings and MOUD prescribing), as has been done previously \cite{cochran2017examination,hoffman2023independent}. 
Further, we dichotomized or otherwise summarized several of the pain-management treatment variables. 
This was necessary to make our analysis tractable, but it likely results in some loss of information and may introduce nondifferential measurement error. Finally, we expect the chronic pain and physical disability subgroups we consider to also be mismeasured, though a previous analysis found the membership in these groups to be relatively stable over time \cite{hoffman2023independent}. 

Another limitation related to our use of claims data is that there is significant attrition of patients over time, 14\% and 21\% of eligible patients were censored by 18- and 24-months respectively. 
We required our cohort to maintain coverage over a 12-month period so that we could characterize our subgroups of interest, covariates, and pain treatments. This means that our inferences apply only to individuals who maintain coverage for at least this period. 
Unlike other previous work, we do not require our cohort to maintain coverage for the entire 24-month study period; instead, we censor them when at the point that they lose Medicaid as their primary insurer. This approach is advantageous because it uses more information and the inferences apply to a larger portion of the Medicaid population. 

In conclusion, we estimated the \textit{unique} contribution of common pain-management treatments to risk of incident OUD in a large, recent cohort of Medicaid patients, disentangling each treatment's contributed risk from the risk conferred by other, potentially co-occurring treatments and confounding factors. Our estimation of the risk associated with 
the co-prescription of gabapentin with opioids is noteworthy both because of its magnitude and 
because of how common this pain treatment is, affecting 9.1\%, and 23.4\% of those in the CP and CPPD subgroups, respectively. 
On a more positive note, our finding that receipt of physical therapy was associated with reduced risk of OUD was also noteworthy, 
because it represents a common and relatively accessible pain-management strategy. 




\vspace{2cm}
\noindent This work was supported by the National Institute on Drug Abuse (grant number R01DA053243). \\


\bibliography{references}

\newpage

\newcommand*{\MyIndent}{\hspace*{0.5cm}}

\singlespacing

\small

\setlength{\LTpost}{0mm}

\newgeometry{left=0.4cm, right=0.4cm, top=2.54cm}

\footnotesize
\begin{longtable}[h]{lcc}
\caption{Descriptive Statistics Stratified by Chronic Pain Subgroup}\\
\hline
\textbf{Characteristic\textsuperscript{1}} & \multicolumn{1}{p{5cm}}{\centering \textbf{Physical Disability \& Chronic Pain} \\ N = 6,133\\ Number (\%)}  & \multicolumn{1}{p{3cm}}{\centering \textbf{Chronic Pain} \\ N = 67,438\\ Number (\%) }\\ 
\midrule
\textbf{Demographics (months 1-6)} &   &   \\ 
\midrule
\textbf{Age} (median, IQR) & 54 (50, 59) &  50 (44, 56)  \\ 
\textbf{Sex} &  &    \\ 
    \MyIndent Female & 3,680 (60) &  36,409 (54) \\ 
    \MyIndent Male & 2,453 (40) &  31,029 (46) \\
\textbf{Race/Ethnicity} &  &    \\ 
    \MyIndent AIAN, non-Hispanic\textsuperscript{2} & 67 (1.2) & 1,270 (2.3) \\ 
    \MyIndent Asian, non-Hispanic & 56 (1.0) &  2,879 (5.2)  \\ 
    \MyIndent Black, non-Hispanic & 1,491 (28) &  7,982 (14)  \\
    \MyIndent Hawaiian/Pacific Islander & \censor{32 (0.6)} & 451 (0.8) \\ 
    \MyIndent Hispanic, all races & 513 (9.5) &  9,221 (17)  \\ 
    \MyIndent Multiracial, non-Hispanic & \censor{3 (<0.1)}  & 52 (<0.1)  \\ 
    \MyIndent White, non-Hispanic & 3,228 (60) &  33,463 (60)  \\  
    \MyIndent Unknown & 743 &  12,120 \\ 
\textbf{Primary Language English} & 4,806 (95) &  52,049 (89)  \\
    \MyIndent Unknown & 1,054 & 8,645  \\ 
\textbf{Married/Partnered} & 502 (13) &  6,746 (26) \\
    \MyIndent Unknown & 2,262 &  41,455  \\ 
\textbf{High Income ($>138$FPL\textsuperscript{2})} & 107 (1.7) & 1,528 (2.3)\\ 
\textbf{Household Size} &  &    \\ 
    \MyIndent 1 & 1,347 (73) & 16,436 (72) \\
    \MyIndent 2 & 266 (14) &3,293 (14)\\ 
    \MyIndent $\ge 3$ & 229 (12) &  3,251 (14) \\ 
    \MyIndent Unknown & 4,291 &  44,458  \\ 
\textbf{Veteran} & \censor{6 (0.7)} &  177 (0.9)  \\ 
    \MyIndent Unknown & 5,250 & 47,819  \\ 
\textbf{TANF Benefits\textsuperscript{2}} & 89 (1.6) & 4,193 (7.6)  \\ 
    \MyIndent Unknown & 650 & 12,315  \\
\textbf{SSI Benefits\textsuperscript{2}} &   &   \\ 
    \MyIndent Mandatory or Optional & 471 (12) & 157 (0.7)  \\ 
    \MyIndent Not Applicable & 3,453 (88) &  22,932 (99)  \\ 
    \MyIndent Unknown & 2,209 & 44,349  \\ 
\midrule
\multicolumn{3}{l}{\textbf{Psychiatric Conditions \& Counseling (months 1-6)}}\\
\midrule
Bipolar & 410 (6.7) &1,399 (2.1)  \\ 
Anxiety & 1,531 (25) & 14,205 (21)  \\ 
ADHD\textsuperscript{2} & 57 (0.9) & 761 (1.1)\\ 
Depression & 1,473 (24) & 9,954 (15)  \\ 
Other Mental Illness & 618 (10) & 4,948 (7.3) \\ 
Mental Health Counseling & 1,006 (16) &  5,460 (8.1)  \\
\midrule
\multicolumn{3}{l}{\textbf{Psychiatric Conditions \& Counseling (months 7-12)}}\\
\midrule
Anxiety & 651 (11) & 5,688 (8.4)  \\ 
Depression & 474 (7.7) &  4,176 (6.2) \\ 
Mental Health Counseling & 1,075 (18) &  6,196 (9.2)  \\
\midrule
\multicolumn{3}{l}{\textbf{Opioid Pain Management (months 7-12)}}\\
\midrule
\textbf{Opioid Pain Prescription} & 3,761 (61) & 27,246 (40) \\ 
    \MyIndent Hydrocodone & 1,599 (26) &  13,334 (20)  \\  
    \MyIndent Oxycodone & 1,790 (29) & 10,049 (15)\\ 
    \MyIndent Fentanyl & 464 (7.6) &  3594 (5.3) \\ 
    \MyIndent Morphine & 419 (6.8) &  2,421 (3.6) \\ 
    \MyIndent Hydromorphone & 294 (4.8) &  2,231 (3.3) \\ 
    \MyIndent Codeine & 283 (4.6) &  2,365 (3.5)\\ 
    \MyIndent Buprenorphine & 27 (0.4) & 170 (0.3)  \\ 
    \MyIndent Other & 75 (1.2)  & 376 (0.6)\\ 
\textbf{Dose, Duration} &  &     \\ 
Max Daily Dose (MME)\textsuperscript{3} & 25 (0, 60) &  0 (0, 36)  \\ 
Proportion of opioid days & 0.05 (0, 0.69) & 0.00 (0.00, 0.08)\\
\textbf{High-Risk Prescribing Practices} &  &     \\ 
Distinct Prescribers\textsuperscript{3} & 1 (0, 2) & 0 (0, 1) \\ 
\textbf{Co-prescription\textsuperscript{3}} & & \\ 
    \MyIndent Benzodiazepine\textsuperscript{3}  & 809 (13) &  3,176 (4.7) \\
    \MyIndent Stimulant\textsuperscript{3} & 58 (0.9) & 476 (0.7) \\ 
    \MyIndent Gabapentinoid\textsuperscript{3} & 1,437 (23) & 6,135 (9.1) \\ 
    \MyIndent Muscle Relaxant\textsuperscript{3} & 1,284 (21) &  6,935 (10) \\
\midrule
\multicolumn{3}{l}{\textbf{Non-opioid Pain Management (months 7-12)}}\\
\midrule
\textbf{Nonopioid Pain Prescription} & 5,213 (85) & 45,250 (67) \\ 
    \MyIndent Antidepressant & 2,973 (49) &  20,846 (31)  \\ 
    \MyIndent Muscle Relaxant & 2,299 (38) &  16,630 (25)  \\ 
    \MyIndent Antiinflammatory \& Antirheumatic & 2,886 (53) & 26,825 (40) \\ 
    \MyIndent Topical & 464 (7.6) &  2,548 (3.8)\\ 
    \MyIndent Other Analgesic \& Antipyretic & 3,382 (55) & 21,541 (32) \\ 
\textbf{Physical Therapy} & 1,581 (26) & 17,932 (27) \\ 
\textbf{Multimodal Pain Treatment} & 1,554 (25) &  17,975 (27)  \\ 
    \MyIndent Ablative Techniques & 82 (1.3) & 646 (1.0) \\ 
    \MyIndent Acupuncture & 15 (0.2) &  506 (0.8) \\ 
    \MyIndent Blocks & 353 (5.8) & 3,436 (5.1) \\ 
    \MyIndent Botulinum Toxin Injection & 59 (1.0) &  436 (0.6) \\
    \MyIndent Chiropractic & 207 (3.4) & 3,204 (4.8) \\ 
    \MyIndent Electrical Nerve Stimulation & 133 (2.2) & 594 (0.9) \\
    \MyIndent Epidural Steroid & 296 (4.8) &  2,796 (4.1)  \\ 
    \MyIndent Intrathecal Drug Therapy & 21 (0.3) &  31 (<0.1)  \\
    \MyIndent Massage Therapy & 725 (12) & 10,221 (15)  \\  
    \MyIndent Trigger Point Injection & 144 (2.3) &  1,235 (1.8)  \\ 
\midrule
\textbf{Outcomes (months 13+)} &  &    \\ 
\midrule
OUD by 18 months\textsuperscript{2} & 410 (7.0) &  1,894 (3.1) \\ 
OUD by 24 months\textsuperscript{2} & 517 (9.7) & 2,280 (4.1)  \\
\bottomrule
\label{tab:desc}
\end{longtable}
\begin{minipage}{\linewidth}
\textsuperscript{1} Median (IQR) for continuous measures; N (\%) for categorical measures \\
\textsuperscript{2} Abbreviations: \\
    \MyIndent AIAN = American Indian and Alaska Native \\
    \MyIndent TANF = Temporary Assistance for Needy Families \\
    \MyIndent SSI = Supplemental Security Income \\
    \MyIndent FPL = Federal Poverty Level \\
    \MyIndent ADD/ADHD = Attention Deficit (Hyperactivity) Disorder \\
    \MyIndent OUD = Opioid Use Disorder \\
\textsuperscript{3} Among patients with any opioid prescription \\
\end{minipage}
\restoregeometry

\renewcommand{\thesection}{S\arabic{section}}
\renewcommand{\theequation}{S\arabic{equation}}

\setcounter{table}{0}
\renewcommand{\thetable}{S\arabic{table}}
\renewcommand\thefigure{S\arabic{figure}}    
\setcounter{figure}{0}    


\maketitle

\section{Additional detail on data and cohort}
We included data from the following states that implemented Medicaid expansion by 2014: ND, VT, NH, CA, OR, MI, IA, NV, OH, IL, NY, MD, MA, RI, HI, WV, WA, KY, DE, AZ, NJ, MN, NM, CT, CO, AR. We excluded MD due to unreliable diagnosis code data\citep{dqatlas}) \cite{KaiserMAP}. We restricted the analysis to expansion states, because they 
nearly all low-income ($\le$138\% of the Federal Poverty Limit) non-elderly adults are eligible for coverage in these states \cite{KaiserMAP}. We permitted enrollment only on or after January 2, 2016, because those with an enrollment date of January 1, 2016 were likely continuously enrolled from 2015, making their initial enrollment date unknown. It was important to know the initial enrollment date, because we limited to the first 24 months of enrollment as after this point, individuals with SSDI would transition to Medicare as the primary payer.

\begin{figure}[H]
\caption{Participant flow diagram for the enrollment cohort used for analyses.}
\includegraphics[width=\textwidth]{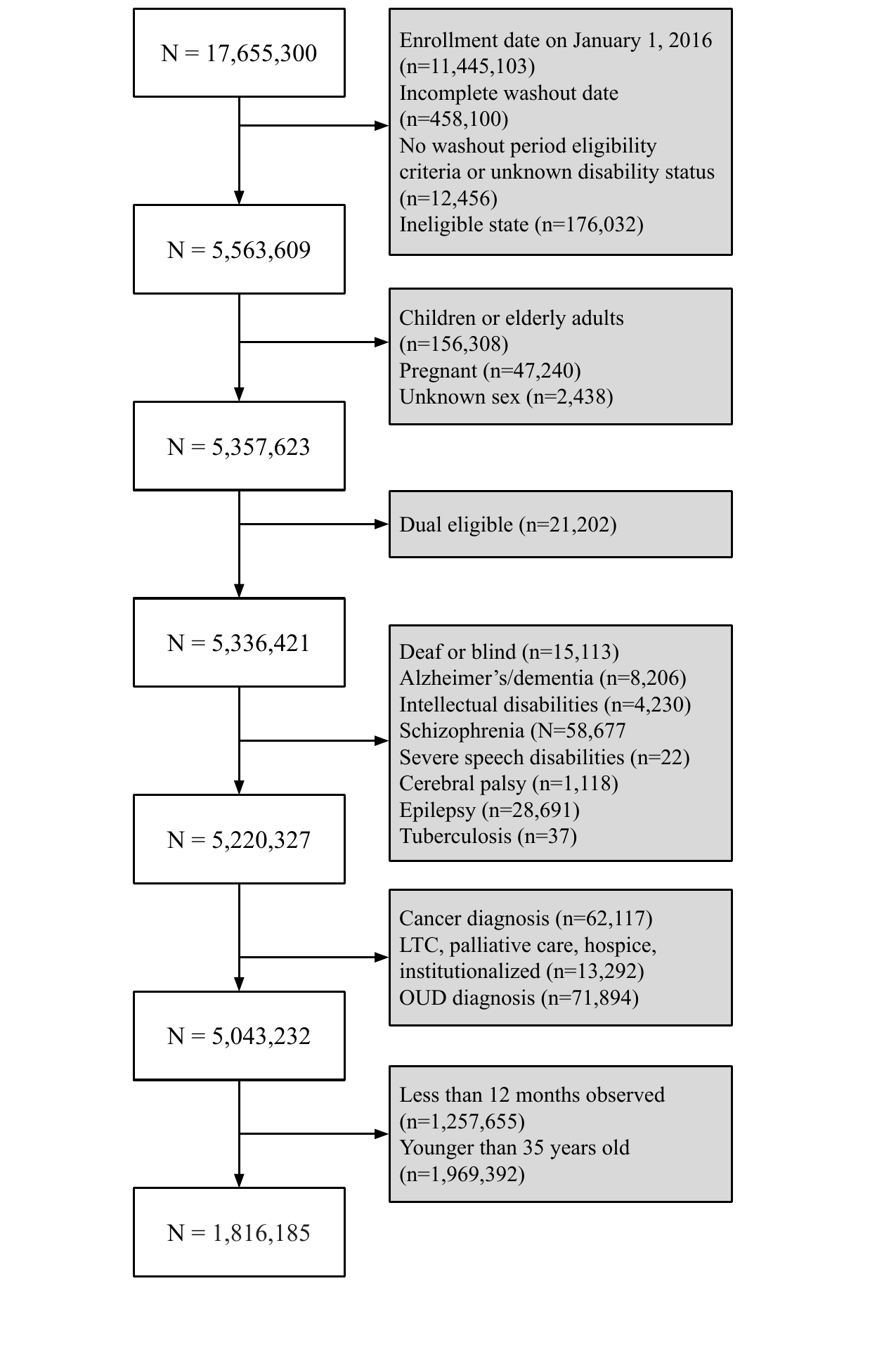}
\label{fig:flowchart}
\end{figure}

\subsection{Additional detail on subgroup definitions}
We had two analytic subgroups defined by chronic pain and physical disability status: 1) co-occurring chronic pain and physical disability (CPPD), and 2) chronic pain without physical disability (CP). Having these subgroups be well-defined is necessary for our research question and causal estimands of interest to also be well-defined, and for understanding to whom our inferences apply. 

 In the pain literature, those with cancer pain are typically excluded for reasons related to the above, as pain treatment priorities would likely be different for this subgroup. Here, we excluded individuals with a cancer diagnosis or in hospice or palliative care from our cohort. 
Identifying beneficiaries with a likely physical disability was impeded by not having access to the beneficiary's qualifying disability coupled with extensive missingness on whether the beneficiary was receiving Medicaid due to SSDI/SSI. Consequently, we used eligibility codes to identify beneficiaries who likely qualified for Medicaid due to a disability, then, used a combination of eligibility codes and exclusion criteria as a means for identifying those with a likely physical disability. Even though we do not explicitly compare effect estimates between subgroups, we applied the following exclusion criteria to the entire cohort to result in more homogeneous, well-defined subgroups who may be ``at risk'' of 
pain management treatments. 
We excluded those who were disabled due to hearing, vision, severe speech, dementia, epilepsy, severe intellectual impairments, 
dementias, intellectual disabilities, cerebral palsy, tuberculosis, epilepsy, and schizophrenia and other serious mental illness (SMI) with psychosis from our cohort.  


\subsection{Additional detail on covariates}
We use the initial 6-month period to characterize the beneficiary's baseline covariates: 
age in years, sex, race/ethnicity (non-Hispanic Asian; non-Hispanic American Indian and Alaska Native (AIAN) or Hawaiian/Pacific Islander; non-Hispanic Black; Hispanic, all races; non-Hispanic White; and non-Hispanic Multiracial or unknown), English as their primary language, marriage/partnership status (currently partnered vs. not), household size (1, 2, and $>2$), veteran status (yes/no), 
income likely $>$133\% FPL (yes/no, determined using the latest washout eligibility code and demographic income information), any inpatient or outpatient DSM-5 diagnosis of: bipolar disorder, any anxiety disorder, attention deficit disorder/ attention deficit hyperactivity disorder (ADD/ADHD), any depressive disorder, or other mental illness (e.g., Anorexia, personality disorders, as operationalized by \citet{samples2018risk}). All relevant ICD codes and operationalizatons are detailed in the Github repository. 

\subsection{Additional detail on statistical analysis}
We assume observed data $\mathbf{O}=(\mathbf{W}, S, \mathbf{A}, \Delta, \Delta Y),$ where: $\mathbf{W}$ represents the covariates measured during months 1-6 (i.e., during the washout period); $S$ represents the chronic condition group (i.e., CPPD or CP), 
characterized during months 1-6; $\mathbf{A}$ represent pain management variables, measured during months 7-12; $\Delta$ represents an indicator of remaining uncensored after the initial enrollment date (24 months for the primary analysis and 18 months for the secondary analysis); finally, $Y$ represents the outcome of incident OUD, which is observed among those who remain uncensored. 

We estimated the effect of each pain management treatment of interest on incident OUD, adjusting for covariates, holding other pain management treatments at their observed levels, 
and stratifying by chronic condition subgroup. Using the notation given above, this effect can be written: $\E(Y^{d_n(\mathbf{A}), \Delta=1} - Y^{\mathbf{A}, \Delta=1} \mid S=s),$ where $\E(Y^{\mathbf{A}, \Delta=1} \mid S=s)$ denotes the expected value of the counterfactual outcome had the set of pain management treatments ($\mathbf{A}$) not been intervened on (i.e., remained as observed) and had no one been censored in subgroup $S=s$, and where $\E(Y^{d_n(\mathbf{A}), \Delta=1} \mid S=s)$ denotes the expected value of the counterfactual outcome had the particular pain management treatment $A_n$ been intervened on as dictated by the function $d_n(\mathbf{A})$ but the remaining pain management treatments stayed as observed and had no one been censored in subgroup $S=s$. 
We considered $N=9$ pain management treatments: $\mathbf{A} = (A_1, ..., A_{9})$. 
For binary pain management treatments (e.g., presence of co-prescription with gabapentin, non-opioid pain prescription, physical therapy, other pain treatment), 
we estimated the effect if everyone in the subgroup received the treatment versus observed treatment. So, for example, if $A_1$ was a binary treatment, we would estimate the effect had everyone received treatment $A_1$, but the remaining pain management treatments remained at their observed levels, i.e.,  $d_1(\mathbf{A}) = (1, A_2, ..., A_{9})$ versus had there been no intervention on the set of treatments, $\mathbf{A}$. 
For non-binary pain management treatments (maximum daily opioid dose, proportion of opioid days, number of opioid prescribers), 
we estimated the effect if everyone in the subgroup had their value increase by 20\% (i.e., multiplied each person's value by 1.2) versus no intervention on treatment. So, for example, if $A_2$ was a non-binary treatment, we would estimate the effect had everyone had 1.2 times their observed value but the remaining pain management treatments remained at their observed levels, i.e., $d_2(\mathbf{A}) = (A_1, 1.2 \times A_2, A_3, ..., A_{9}).$ 

We consider causal effects that isolate the effect of each treatment variable (holding the other treatment variables at their observed levels and adjusting for confounding and censoring) on the outcome under identifying assumptions of: 1) conditional exchangeability, meaning that there is no unobserved/unmeasured confounding of the relationship between the set of treatments 
and outcome 
conditional on covariates and that there is no unobserved/unmeasured confounding between censoring and the outcome conditional on the covariates and treatment; and 2) positivity, meaning that if it is possible to find a beneficiary with observed data $(\mathbf{a},c, \mathbf{w})$, then it is possible to find a beneficiary with data $(d_n(\mathbf{a}), 1,\mathbf{w})$ for $n \in \{1:9\}$. In other words, this states that: 1) if it is possible to find a beneficiary with covariates $\mathbf{w}$ and treatment set $\mathbf{a}$, it is possible to find a beneficiary with covariates $\mathbf{w}$ and treatment set $d_n(\mathbf{a})$ for $n \in \{1:9\}$, and 2) for every observed $(\mathbf{a}, \mathbf{w})$, there is a nonzero probability of not being censored. 

\begin{figure}[ht]
\caption{Estimated effects of each pain management treatment on incident OUD, as defined by ICD-10 diagnosis codes, and 95\% CIs for the cohort with (a) disability and co-occurring chronic pain and (b) chronic pain alone.}
\centering
 \subfloat[Disability and co-occurring chronic pain \label{fig:pdcpicd}]{
\includegraphics[width=\textwidth]{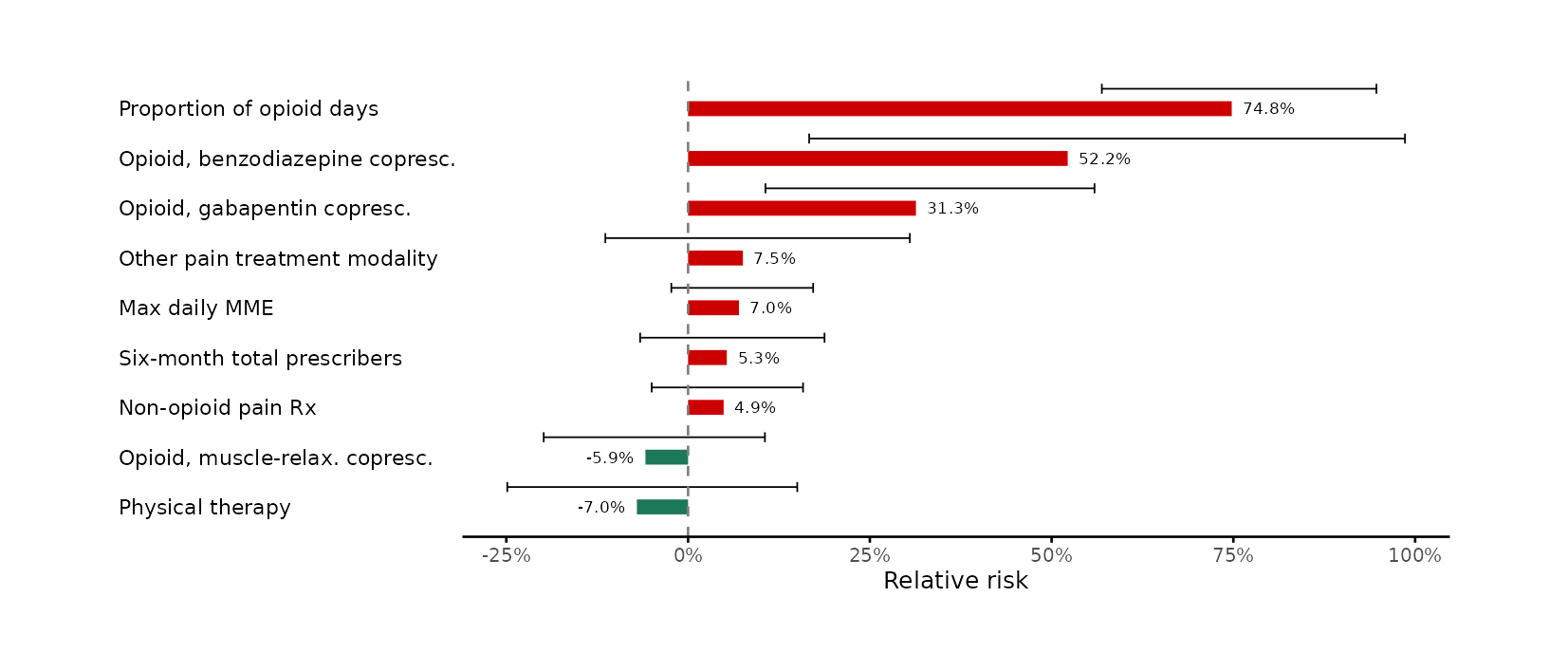}
 }

 \vspace{-.3cm}
  \subfloat[Chronic pain alone \label{fig:cpicd}]{
\includegraphics[width=\textwidth]{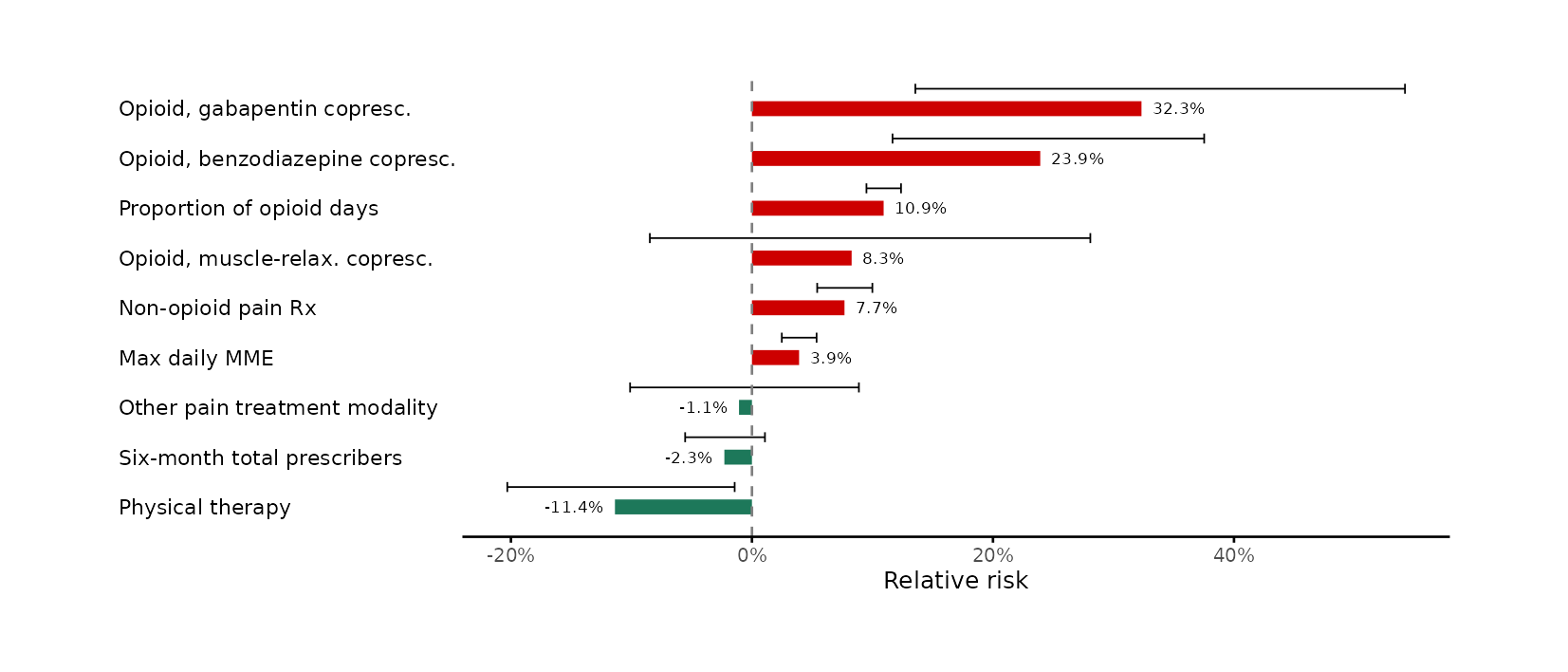}
 }
   \label{fig:resicd}
   
\end{figure}

\begin{figure}[ht]
\caption{Estimated effects of each pain management treatment on incident OUD by 6 months of follow up, as defined by the expanded definition used in the primary outcome, and 95\% CIs for the cohort with (a) disability and co-occurring chronic pain and (b) chronic pain alone.}
\centering
 \subfloat[Disability and co-occurring chronic pain \label{fig:pdcp18}]{
\includegraphics[width=\textwidth]{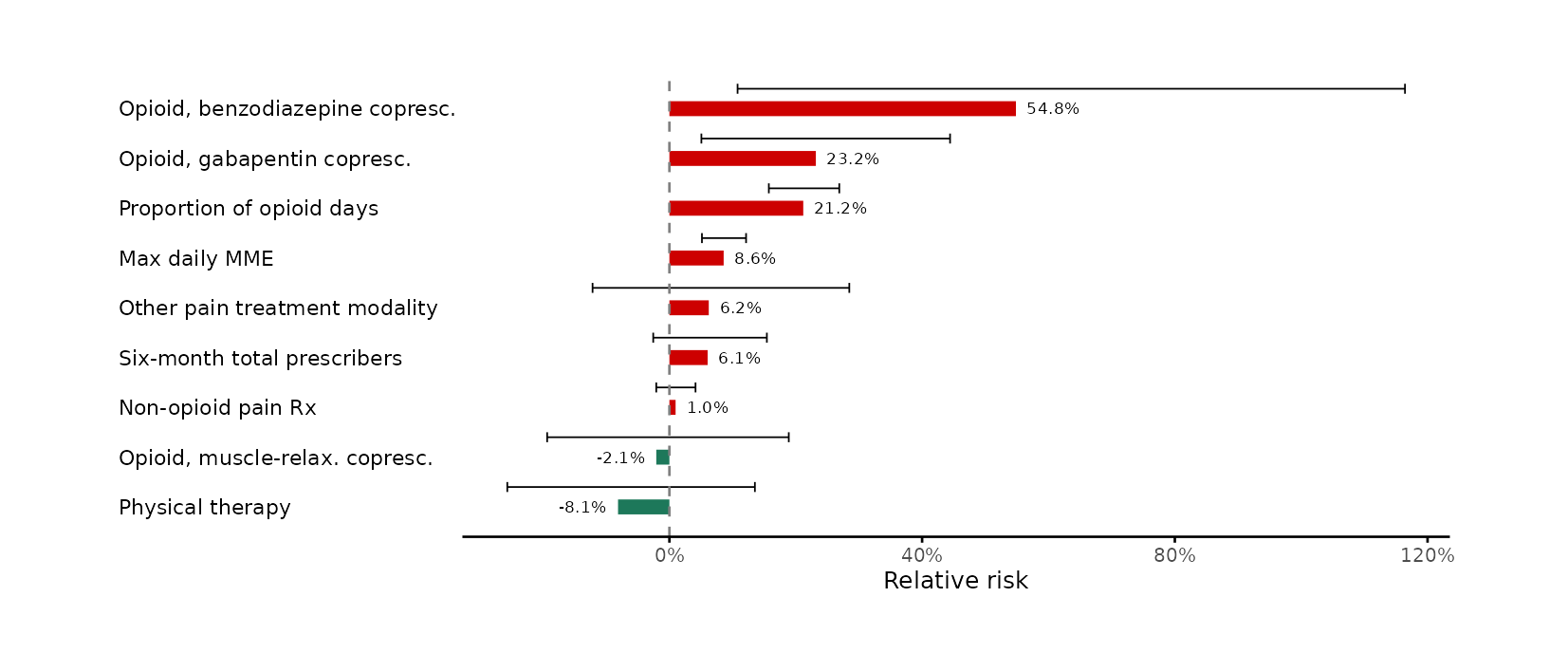}
 }
 \vspace{-.3cm}
  \subfloat[Chronic pain alone \label{fig:cp18}]{
\includegraphics[width=\textwidth]{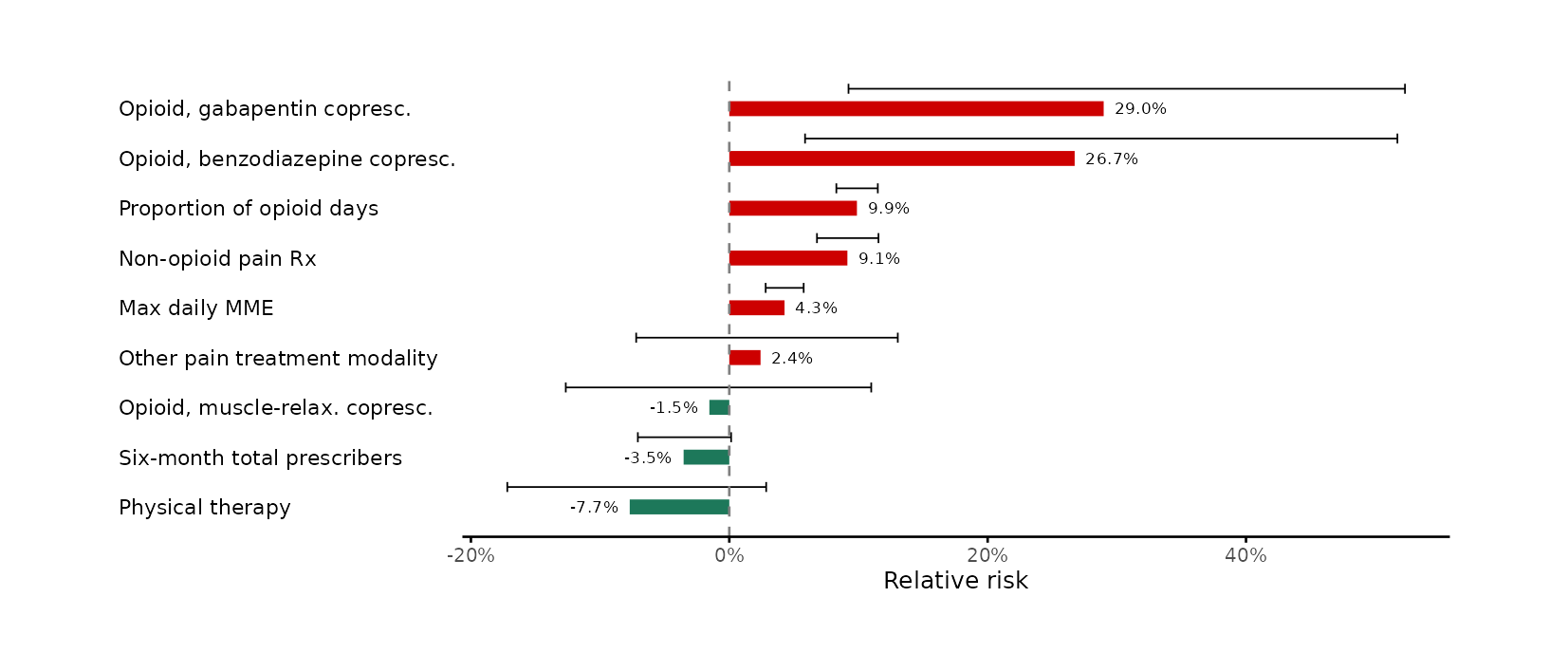}
 }
   \label{fig:res18}
   
\end{figure}

\end{document}